\documentclass[%
 reprint,
 amsmath,amssymb,
 aps,
 prl,
]{revtex4-1}

\usepackage{mathrsfs}
\usepackage{graphicx}
\usepackage{dcolumn}
\usepackage{bm}
\usepackage{amsmath,amssymb}
\usepackage[caption=false]{subfig}
\usepackage{bm,wasysym,color}

\begin{document}

\title{Electrically Tunable Optical Nonlinearities in Graphene-Covered SiN Waveguides Characterized by Four-Wave Mixing} 

\author{Koen Alexander}
\email{Koen.Alexander@UGent.be}
\author{Bart Kuyken}
\author{Dries Van Thourhout}
\affiliation{
Photonics Research Group, INTEC, Ghent University-IMEC, Ghent B-9000, Belgium
}
\affiliation{
Center for Nano-and Biophotonics (NB-Photonics), Ghent University, Ghent B-9000, Belgium
}
\author{N. A. Savostianova, S. A. Mikhailov}
\affiliation{
 Institute of Physics, University of Augsburg, D-86135 Augsburg, Germany
}

\date{\today}

\begin{abstract}
We present a degenerate four-wave mixing experiment on a silicon nitride (SiN) waveguide covered with gated graphene. We observe strong dependencies on signal-pump detuning and Fermi energy, i.e. the optical nonlinearity is demonstrated to be electrically tunable. In the vicinity of the interband absorption edge ($2|E_F|\approx \hbar\omega$) a peak value of the waveguide nonlinear parameter of $\approx$ 6400 m$^{-1}$W$^{-1}$, corresponding to a graphene nonlinear sheet conductivity $|\sigma_s^{(3)}|\approx4.3\cdot 10^{-19}$ A m$^2$V$^{-3}$ is measured.
\end{abstract}

\maketitle

In recent years, there has been increasing interest in the nonlinear optical properties of graphene. Both theoretical predictions \cite{mikhailov2007non,Mikhailov2008, cheng2014, cheng2015, Mikhailov2016, semnani2016} and experimental studies \cite{Hendry2010, Zhang2012, chen2013, Miao2015, Dremetsika2016} have indicated that graphene has a very high third-order sheet conductivity $\sigma_s^{(3)}$, which leads to a strong nonlinear optical response. Despite the consensus that nonlinearities in graphene are strong, very different values of the corresponding material parameters have been reported (see, e.g., discussions in Refs. \cite{cheng2014, Dremetsika2016}). Reasons for this can be found in the fact that in different experiments different nonlinear effects (harmonics generation, four-wave mixing, Kerr effect) are probed at different wavelengths, in samples with different carrier densities and in different dielectric environments. Moreover, in experimental studies pulses with vastly different durations and optical bandwidths have been used. All these factors can significantly influence the final result. Hence a detailed quantitative study of the nonlinear response of graphene, at different frequencies and in samples with different electron densities, is imperative.

\begin{figure*}
    \includegraphics[width=\textwidth]{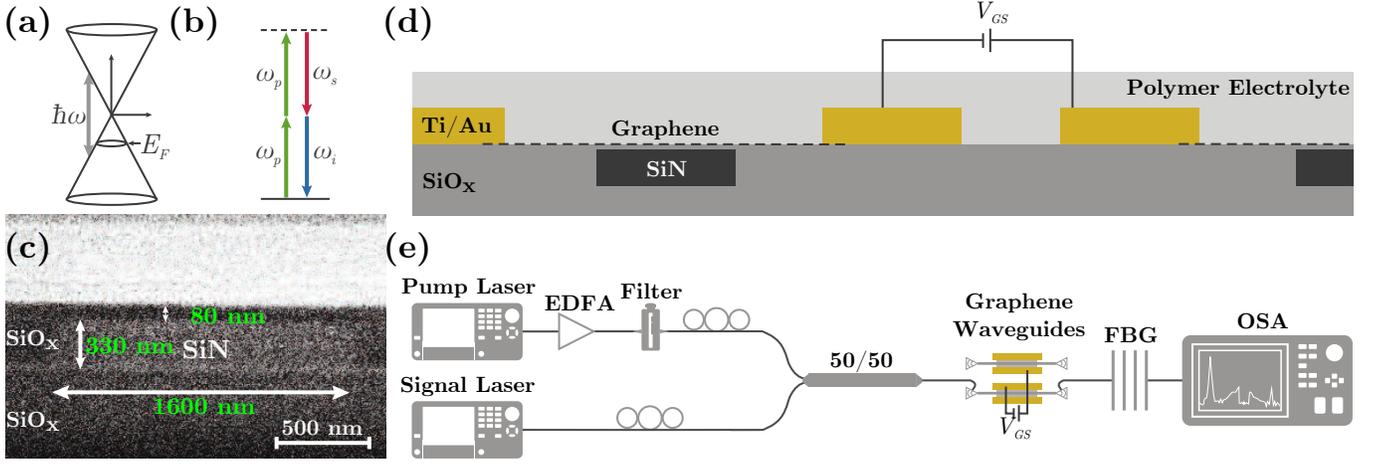}
    \caption{\label{fig:sample_setup}a) Graphene band diagram. b) Degenerate FWM energy diagram. c) SEM image of the cross-section of a SiN waveguide. d) Sketch of the gating scheme. e) Setup used for the FWM experiments.}
\end{figure*}

Another important research direction is the search for interband resonances in the nonlinear response function of graphene. It has been theoretically predicted \cite{cheng2014, cheng2015, semnani2016, Mikhailov2016} that the nonlinear parameters of graphene should have resonances at frequencies corresponding to the interband absorption edge, e.g. at $\hbar\omega=2|E_F|/3$ for third harmonic generation or at $\hbar\omega=2|E_F|$ for self-phase modulation (SPM) (SPM in graphene with fixed $E_F$ on a waveguide has been measured e.g. in Ref. \cite{vermeulen2016negative}), etc., where $\omega$ is the incident photon frequency and $E_F$ the Fermi energy. Fig. \ref{fig:sample_setup}a shows the band diagram of graphene, along with the photon energy. SPM, which scales as Im$[\sigma_s^{(3)}]$, has been predicted to peak at $2|E_F|\approx \hbar \omega$, after which it decreases sharply for further increasing $|E_F|$ (see for example Fig. 3 in Ref. \cite{vermeulen2016opportunities}). The majority of experiments have been performed in weakly doped graphene at high frequencies (near-IR, visible), where $\hbar\omega\gg 2|E_F|$. An experimental observation of the Fermi energy related resonances would not only be interesting for fundamental science but also for nonlinear devices since it would provide a way to control the nonlinear optical response of practical systems.

In this Letter, we characterize the Fermi energy dependence of the third order nonlinear effects in graphene, tuning the graphene from intrinsic ($2|E_F|\ll\hbar\omega$) to beyond the interband absorption edge ($2|E_F|>\hbar\omega$). We do this by means of four-wave mixing (FWM) in an integrated silicon nitride (SiN) waveguide, covered with a monolayer of graphene. Because of the coupling between the evanescent tail of the highly confined waveguide mode and the graphene over a relatively long length, significant light-matter interaction can be achieved. Studies of nonlinear effects in graphene-covered silicon waveguides and resonators have been published previously \cite{gu2012regenerative, zhou2014enhanced, ji2015enhanced, vermeulen2016negative}. However, an intrinsic disadvantage of using a silicon platform for the characterization of graphene nonlinearities is that silicon has a relatively strong nonlinear response itself. The real part of the nonlinear parameter of a typical Si waveguide is about $\gamma_{Si} \approx 300$ m$^{-1}$W$^{-1}$ \cite{moss2013new}, as opposed to about $\gamma_{SiN} \approx 1.4$ m$^{-1}$W$^{-1}$ \cite{moss2013new} for a SiN waveguide, which is negligible compared to the nonlinear parameters of the graphene-covered waveguide measured in this work ($\gamma P L$ is the nonlinear phase shift acquired over length $L$ at power $P$, see Supplemental Material or Ref. \cite{agrawal2007nonlinear}). Using SiN, we can thus avoid any ambiguity about the origin of the strong nonlinear effects. Furthermore, we have achieved electrical tuning of $E_F$ (gating) by using a polymer electrolyte \cite{das2008monitoring}. We have performed measurements for a varying signal-pump detuning and for a broad range of charge carrier densities and demonstrate, for the first time to our knowledge, a significant increase of the nonlinear response of graphene in the vicinity the interband absorption edge $\hbar\omega\approx2|E_F|$. Moreover, we demonstrate a good qualitative agreement with theoretical calculations.

Degenerate four-wave mixing (FWM) is a third order nonlinear optical process in which two pump photons at frequency $\omega_p$ are converted into two photons at different frequencies, typically denoted as the signal $\omega_s$ and idler $\omega_i$. Energy conservation dictates that $\omega_s+\omega_i=2\omega_p$, which is schematically shown in Fig. \ref{fig:sample_setup}b. For a small signal-pump detuning, $\omega_p\approx\omega_s\approx\omega_i$, the theory  predicts a vanishing FWM response beyond the interband absorption edge, $2|E_F|>\hbar\omega_p$, in analogy with SPM described before. However, as opposed to SPM, the theory as it is published in Refs. \cite{cheng2015, Mikhailov2016} does not predict a sharp peak at $2|E_F|\approx\hbar\omega_p$. This is because the FWM response scales as the absolute value of the third order conductivity, $|\sigma_s^{(3)}|$, in which Re$[\sigma_s^{(3)}]$ typically dominates Im$[\sigma_s^{(3)}]$.

To investigate this experimentally, a pump at a fixed wavelength and a signal with variable wavelength are injected into the graphene-covered SiN waveguide. Under the current experimental conditions, one can prove that the conversion efficiency $\eta$, defined as the ratio between the idler power to the signal power at the output, is quadratically dependent on the nonlinear parameter $\gamma$ of the waveguide (see Supplemental Material),
\begin{align}
& \eta\equiv\frac{P_i(L)}{P_s(L)}\approx\vert \gamma(\omega_i;\omega_p,\omega_p,-\omega_s) \vert^2 \left(\frac{\omega_i}{\omega_p} \right)^2 P_p(0)^2L_{\text{eff}}^2,
\label{eq:eta_main}
\end{align}
where $L$ is the length of the graphene-covered waveguide section, $L_{\text{eff}}\approx \frac{1-e^{-\alpha L}}{\alpha}$ the effective length of the nonlinear process and $\alpha$ the linear waveguide loss. The effect of the phase mismatch is neglected since $L\beta_2\Delta\omega^2 \ll 1$ in the presented experiment ($L=100\;\mu$m, $\Delta\omega<10^{13}$ rad/s and $\beta_2$ of a SiN waveguide is on the order of $10^{-25}$ s$^2$/m \cite{moss2013new}; here $\beta_2\equiv \frac{\partial^2 \beta}{\partial \omega ^2 }$, with $\beta(\omega)$ the propagation constant of the optical mode). The nonlinear parameter $\gamma$ of the waveguide is, to a good approximation, proportional to the nonlinear conductivity $\sigma_s^{(3)}$ of graphene (see Supplemental Material),
\begin{align}
\begin{split}
&\gamma(\omega_i;\omega_p,\omega_p,-\omega_s)\approx\\
&i \frac{3 \sigma^{(3)}_{s,\;xxxx}(\omega_i; \omega_p, \omega_p, -\omega_s)}{16\mathscr{P}_p^2}\int_{G}\vert\mathbf{e}(\omega_p)_\parallel\times \mathbf{\hat{e}}_z\vert^4  d\ell\;,
\end{split}
\label{eq:gamma_main}
\end{align}
where $\mathbf{e}(\omega_p)_\parallel$ is the electric field component tangential to the graphene sheet at the pump frequency, $\mathbf{\hat{e}}_z$ is the unit vector along the propagation direction and $\mathscr{P}_p$ is the power normalization constant of the optical mode.

A set of straight waveguides was patterned in a 330 nm thick LPCVD SiN layer on top of a 3 $\mu$m burried oxide layer on a silicon handle wafer. The sample was then covered with LPCVD oxide and planarized using a combination of chemical mechanical polishing, reactive ion etching and wet etching. Subsequently, a CVD-grown graphene layer was transferred to the samples by Graphenea \cite{Graphenea2017} and patterned using photolithography and oxygen plasma etching so that different waveguides were covered with different lengths of graphene. Metallic contacts (Ti/Au; $\approx$ 5 nm/300 nm) were applied at both sides of each waveguide, with a spacing of 12 $\mu$m. Fig. \ref{fig:sample_setup}c shows a SEM image of the waveguide cross-section, note that $\approx$ 80 nm of oxide is left on the waveguide. Finally the structures were covered with a polymer electrolyte consisting of LiClO$_4$ and polyethylene oxide (PEO) in a weight ratio of 0.1:1. Fig. \ref{fig:sample_setup}d shows a sketch of the cross-section (not to scale). The gate voltage $V_{GS}$ can be used to gate the graphene layer \cite{das2008monitoring}. The dependence of $E_F$ on $V_{GS}$ can be approximated by the following formula \cite{thareja2015electrically,das2008monitoring}:
\begin{align}
V_{GS}-V_D = \mathrm{sgn}(E_F)\frac{e E_F^2}{\hbar^2 v_F^2 \pi C_{EDL}}+\frac{E_F}{e} \label{eq:VvsEf_main}\;,
\end{align}
where $e$ is the electron charge, $v_F\approx 10^6$ m/s the Fermi velocity, $C_{EDL}$ the electric double layer capacitance and $V_D$ the Dirac voltage. Based on measurements of the optical loss and the graphene sheet resistance versus the gate voltage we estimated  $C_{EDL}\approx1.8\cdot10^{-2}$ F m$^{-2}$ and $V_D\approx0.64$ V, see Supplemental Material.

The setup used for the FWM experiment is shown in Fig. \ref{fig:sample_setup}e. A pump laser (Syntune S7500, $\lambda_p=$ 1550.18 nm) is amplified using an Erbium-doped fiber amplifier (EDFA), a tunable band-pass filter 
suppresses the Amplified Spontaneous Emission (ASE) of the EDFA. The signal is provided by a Santec Tunable Laser TSL-510. Pump and signal are coupled into the waveguide through a grating coupler. At the output a fiber Bragg grating (FBG) filters out the strong pump light and the signal and idler are visualized on an Anritsu MS9740A optical spectrum analyser (OSA).
\begin{figure}[ht]
	\centering
	\subfloat{\includegraphics{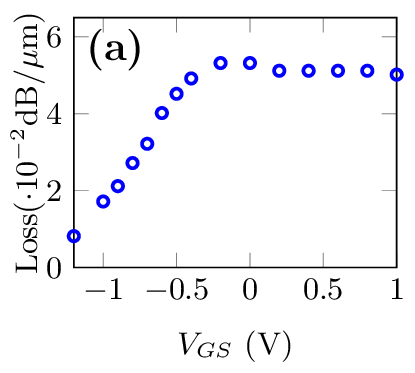}\label{fig:loss}}
	\subfloat{\includegraphics{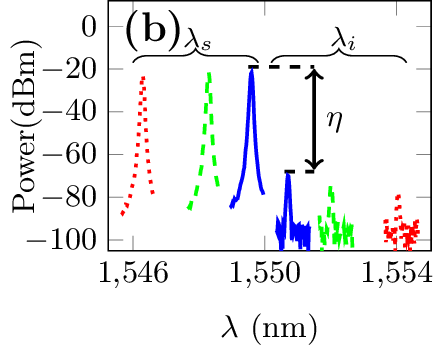}\label{fig:example_spectra}}\\[-12pt]
	\subfloat{\includegraphics{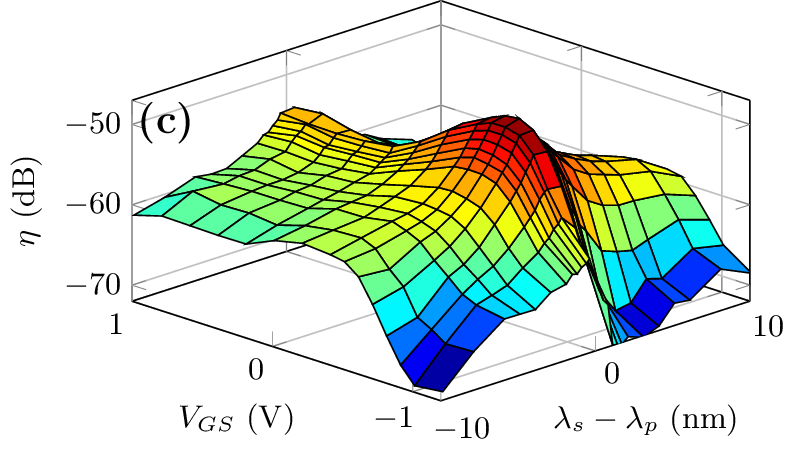}\label{fig:eta_surf}}
	\caption{a) Waveguide loss as a function of gating voltage. b) Examples of the optical spectra ($V_{GS}=-0.5$ V). The pump peak (1550.18 nm) is filtered out by the FBG. The signal peaks can be seen on the left and the corresponding idler peaks on the right. Graphene section length: 100 $\mu$m. c) Conversion efficiency $\eta$ as a function of $V_{GS}$ and detuning $\lambda_s-\lambda_p$.\label{fig:all_measurements}} 
\end{figure}
\begin{figure*}[ht]	
	\centering
	\subfloat{\includegraphics{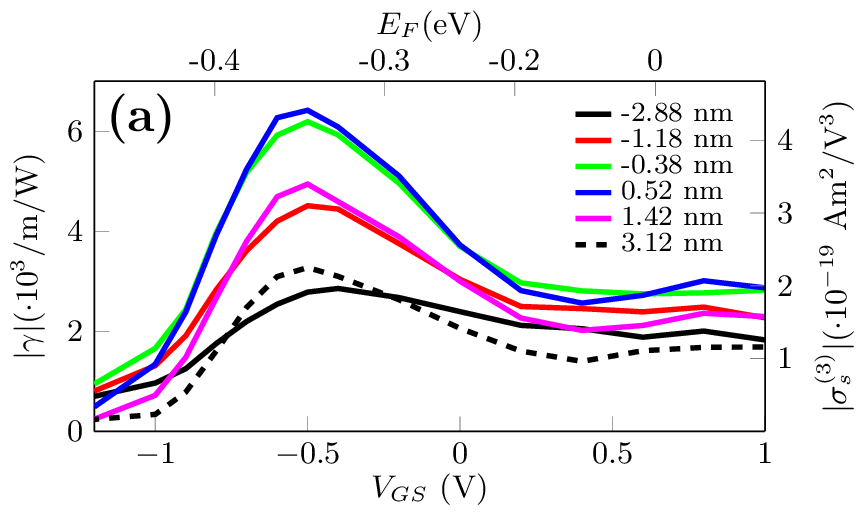}\label{fig:gamma_V}}
	\subfloat{\includegraphics{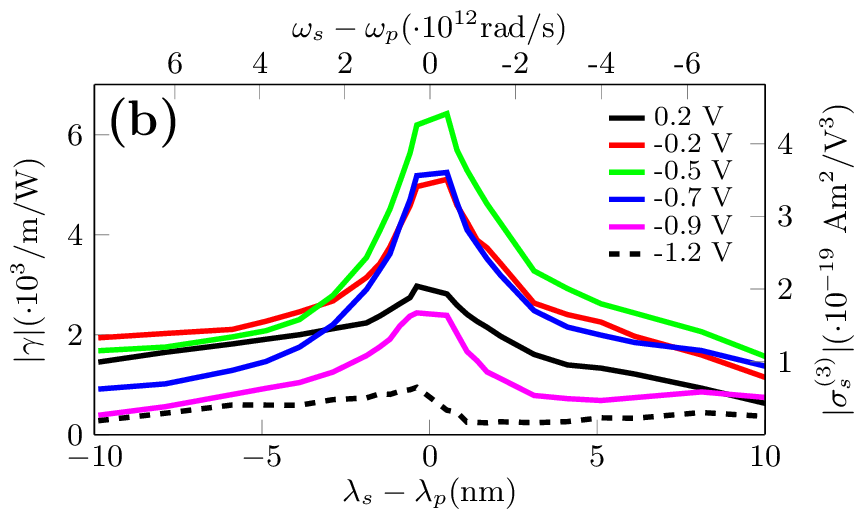}\label{fig:gamma_detuning}}\\[-4pt]
	\hspace{-31pt}
	\subfloat{\includegraphics{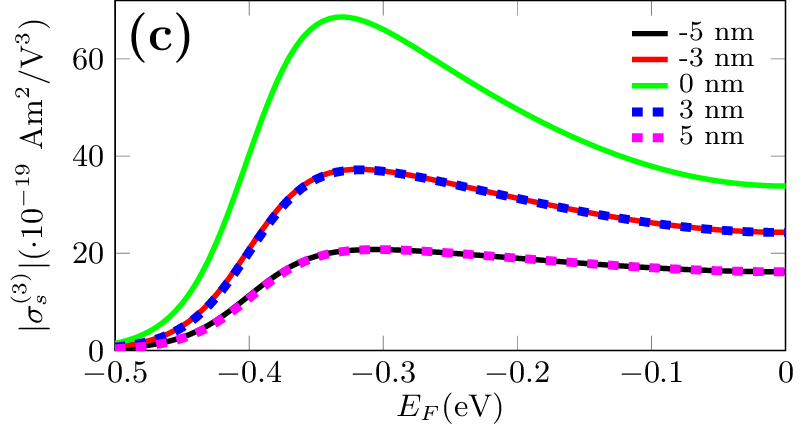}\label{fig:gamma_Ef_simulation}}
	\hspace{14pt}
	\subfloat{\includegraphics{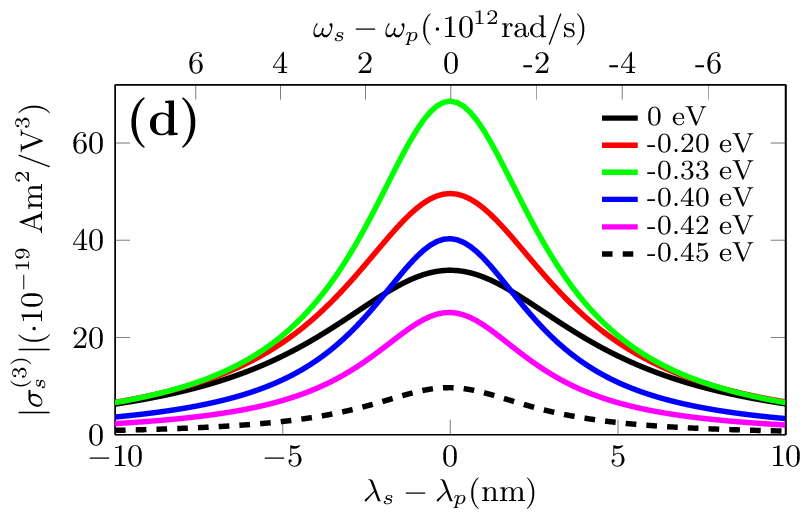}\label{fig:gamma_detuning_simulation}}
	\caption{(a) and (b): measured nonlinear parameter $|\gamma(\omega_i;\omega_p,\omega_p,-\omega_s)|$/graphene nonlinear conductivity $|\sigma^{(3)}_{s}(\omega_i; \omega_p, \omega_p, -\omega_s)|$. As a function of gate voltage/Fermi energy for different values of detuning $\lambda_s-\lambda_p$ (a). As a function of detuning for different values of the gate voltage (b). (c) and (d): calculated values of $|\sigma^{(3)}_{s}(\omega_i; \omega_p, \omega_p, -\omega_s)|$, as a function of Fermi energy for different wavelength detunings (c), and as function of detuning for a range of Fermi energies (d).\label{fig:gamma}} 
\end{figure*}

Fig. \ref{fig:all_measurements} summarizes the experimental results obtained for a set of 1600 nm wide waveguides. By measuring the transmission of a set of waveguides with varying graphene lengths, as well as the transmission as a function of $V_{GS}$, the propagation loss as a function of $V_{GS}$ was extrapolated, Fig. \ref{fig:loss}. Note that the absorption drops sharply for negative voltages, indicating that the Fermi level of the graphene gets tuned beyond the interband absorption edge. In Fig. \ref{fig:example_spectra}, some of the measured spectra  for the FWM experiment are plotted ($V_{GS}=-0.5$ V), the pump laser light is filtered out by the FBG. The conversion efficiency $\eta$ can be read as the ratio between the idler ($\lambda_i$) and signal ($\lambda_s$) peaks (we correct for variations in the transmission of the grating couplers with changing wavelength). The measured conversion efficiencies are plotted in Fig. \ref{fig:eta_surf}, for a range of different voltages and signal wavelengths ($\lambda_p=1550.18$ nm), with an estimated on-chip pump power of $P_p(0)=10.5$ dBm and a graphene length $L=100\;\mu$m. The FWM conversion efficiency is highly dependent on both detuning $\lambda_s-\lambda_p$ and the applied voltage. 

Using Eqs. \eqref{eq:eta_main} and \eqref{eq:gamma_main}, the magnitude of the nonlinear parameter $\gamma(\omega_i;\omega_p,\omega_p,-\omega_s)$ and of the third order conductivity $\sigma^{(3)}_{s}(\omega_i; \omega_p, \omega_p, -\omega_s)$ can be calculated. For the calculation of $L_{\text{eff}}$ the loss measurement in Fig. \ref{fig:loss} was used. The integral and power normalization constant $\mathscr{P}_p$ in Eq. \eqref{eq:gamma_main} are calculated using a COMSOL Multiphysics$^\circledR$-model of the cross-section in Fig. \ref{fig:sample_setup}c. Figs. \ref{fig:gamma}a and \ref{fig:gamma}b show the results of these conversions. The measured values for $|\gamma|$ ($|\sigma^{(3)}_{s}|$) have a sharp resonance as a function of detuning and a broad asymmetric resonance as a function of $E_F$. $|\gamma|$ ($|\sigma^{(3)}_{s}|$) is about 2800 m$^{-1}$W$^{-1}$ ($2\cdot10^{-19}$ Am$^2$/V$^3$) at small $|E_F|$ (for minimum detuning) and about 6400 m$^{-1}$W$^{-1}$($4.3\cdot10^{-19}$ Am$^2$/V$^3$) at its absolute peak.

We can compare these experimental results with a slightly modified version of the theory published in Refs. \cite{Mikhailov2016, cheng2015}. In these papers analytical expressions for the third order conductivity $\sigma^{(3)}_{s,\;\alpha\beta\gamma\delta}(\omega_1+\omega_2+\omega_3;\omega_1,\omega_2,\omega_3,E_F,\Gamma)$ were derived at $T=0$, where the relaxation rate $\Gamma$ was assumed to be energy independent. However, under these assumptions the theory does not predict the measured increase of $\sigma^{(3)}_{s}$ with increasing $|E_F|$ (see the solid line on Fig. S5 in the Supplemental Material). To get a better correspondence between theory and experiment, we can however assume that $\Gamma(E)$ is a function of the electron energy. Both theoretical (e.g. Ref. \cite{sarma2011electronic}) and experimental (e.g. Ref. \cite{tan2007measurement}) studies indicate that the relaxation rate $\Gamma(E)\propto |E|^{-\alpha}$ is a power-law function of energy $E$, at $|E|\gtrsim E_0$, with $\alpha$ being determined by the scattering mechanism. According to theory, $\alpha=1$ for impurity scattering and $E_0$ is related to the density of impurities \cite{sarma2011electronic}. Experimental data confirmed the power-law dependence of $\Gamma(E)$ but showed a slightly smaller value of $\alpha$, $0.5\lesssim \alpha\lesssim 1$ (see the inset in Fig. 2 in Ref. \cite{tan2007measurement}). To be able to use a formula for $\Gamma(E)$ at all energies including the limit $E\to 0$ we adopt the model  
\begin{align}
\Gamma(E)=\frac{\Gamma_0}{\left(1+E^2/E_0^2\right)^{\alpha/2}},
\label{gammaE}
\end{align}
which has a correct asymptote $\Gamma(E)\propto |E|^{-\alpha}$ at large energies $|E|\gg E_0$ and gives a constant relaxation rate at $E\to 0$; the quantities $\Gamma_0$, $E_0$ and $\alpha$ in Eq. (\ref{gammaE}) are treated as fitting parameters. In addition, to take into account the effects of nonzero temperatures, we can use the formula (the frequency arguments are omitted for clarity) \cite{cheng2015}:
\begin{align} 
\begin{split}
&\sigma^{(3)}_{s,\;\alpha\beta\gamma\delta} (E_F, \Gamma_0,E_0,\alpha,T)=\\
&\frac 1{4T}\int_{-\infty}^{+\infty}\frac{\sigma^{(3)}_{s,\;\alpha\beta\gamma\delta} (E_F',\Gamma_0,E_0,\alpha,T=0) }{\cosh^2\left(\frac{E_F-E_F'}{2T}\right)} dE_F'.
\end{split}
\end{align}
Figs. \ref{fig:gamma}c and \ref{fig:gamma}d show thus obtained theoretical dependencies of the absolute value of the third order conductivity $|\sigma^{(3)}_{s,\;xxxx}(\omega_i;\omega_p,\omega_p,-\omega_s)|$ on the Fermi energy and the detuning $\lambda_s-\lambda_p$. The parameters $\hbar\Gamma_0=2.5$ meV, $E_0=250$ meV and $\alpha=0.8$ have been chosen so that good qualitative agreement was obtained with the experimental plots shown in Figs. \ref{fig:gamma}a and \ref{fig:gamma}b. One can see that the theory indeed describes the most important features of the FWM response: a narrow resonance as a function of $\lambda_s-\lambda_p$ and a broad strongly asymmetric shape as a function of $E_F$; the inflection point at $E_F\approx -0.4$ eV corresponds to $\hbar\omega_p\approx 2|E_F|$. Quantitatively, the theory predicts about one order of magnitude \textit{larger} response than was experimentally observed (this contrasts to a number of previous publications, where the measured nonlinear response was claimed to be larger than the theoretically calculated one, see discussions in Refs. \cite{cheng2014, cheng2015, vermeulen2016opportunities}). This discrepancy should be subject to further investigation. Experimental errors could have an influence, such as an overestimated pump power $P_p(0)$ or effective length $L_{\text{eff}}$, errors on the exact dimensions of the waveguide cross-section, etc. Inhomogeneities in the doping level of the graphene could also have an influence, effectively creating inhomogeneous broadening of the measured response in the $|E_F|$-direction and diminishing the height of the peak. Finally, the difference could partly be due to imperfections in the graphene which the theory might fail to fully take into account.

In conclusion, we have performed a degenerate four-wave mixing experiment on a graphene-covered SiN waveguide. A polymer electrolyte enabled us to gate the graphene over a relatively large window. The experiment shows that the nonlinear conductivity of graphene has a sharp resonance as a function of signal-pump detuning, also a broad asymmetric resonance shape in the vicinity of the absorption edge $2|E_F|=\hbar \omega_p$ is observed. Qualitative agreement was obtained between these experimental data and an adapted version of previously published theory \cite{Mikhailov2016, cheng2015}, in which we introduced an energy-dependent relaxation rate $\Gamma(E)$. From an application perspective, it is important to note that the measured nonlinear parameter of the waveguide $|\gamma|$ is tunable by applying a gate voltage, and that optimizing this voltage the parameter surpasses $\approx2000$ m$^{-1}$W$^{-1}$ over the full measured bandwidth of 20 nm, with peak values over $\approx6000$ m$^{-1}$W$^{-1}$. This is more than 3 orders of magnitude larger than the nonlinear parameter of a standard SiN waveguide. The most obvious trade-off is the strongly increased linear absorption (more than 2 orders of magnitude, though this absorption is also tunable with voltage). The question of whether a graphene-covered integrated waveguide platform could serve for actual applications requires further research, such as experiments that quantify not only the magnitude, but also the phase of $\gamma$, experiments to test how the material behaves over larger bandwidths and at higher optical powers, optimizations of the waveguide cross-section, etc. The quantitative gap between theoretical and experimental results is another issue that needs further investigation.

We thank Prof. Daniel Neumaier and Dr. Muhammad Mohsin for useful discussions and for providing the polymer electrolyte. We also thank Dr. Owen Marshall for giving useful advice. The work has received funding from the European Union’s Horizon 2020 research and innovation programme GrapheneCore1 under Grant Agreement No. 696656. K. A. is funded by FWO Flanders.

\pagebreak
\widetext
\begin{center}
\textbf{\Large Supplemental Material}
\end{center}
\setcounter{equation}{0}
\setcounter{figure}{0}
\setcounter{table}{0}
\setcounter{page}{1}
\makeatletter
\renewcommand{\theequation}{S\arabic{equation}}
\renewcommand{\thefigure}{S\arabic{figure}}
\renewcommand{\bibnumfmt}[1]{[S#1]}
\renewcommand{\citenumfont}[1]{S#1}
\fontsize{12}{17}\selectfont

\section{Theory  of four-wave mixing in graphene-covered SiN waveguides} 
In this section the expressions for the nonlinear parameters and the linear loss of a waveguide covered with graphene are derived. In this derivation, the nonlinearities and linear losses will be treated as perturbations to the ideal lossless, linear waveguide. In addition, an expression for the degenerate four-wave mixing conversion efficiency $\eta$ is derived.

The complex amplitude of an unperturbed waveguide mode at frequency $\omega_j$ can be written as:
\begin{align}
\mathbf{E}_{0}(\omega_j,\mathbf{r})=& A_0\frac{\mathbf{e}(\omega_j,\mathbf{r}_\perp)}{\sqrt{\mathscr{P}_j}}e^{i\beta_j z} \;,\label{eq:EUnperturbed}\\
\mathbf{H}_{0}(\omega_j,\mathbf{r})=& A_0\frac{\mathbf{h}(\omega_j,\mathbf{r}_\perp)}{\sqrt{\mathscr{P}_j}}e^{i\beta_j z} \;. \label{eq:HUnperturbed}
\end{align}
Where $\mathbf{e}(\omega_j,\mathbf{r}_\perp)$ and $\mathbf{h}(\omega_j,\mathbf{r}_\perp)$ are the vectorial electric and magnetic mode profiles, in what follows, we will often omit the arguments $\mathbf{r}$ and $\mathbf{r}_\perp$ for brevity. $A_0$ is the complex amplitude of the mode. $\beta_j$ is the mode propagation constant and $\mathscr{P}_j$ is the power normalization constant, defined so that the total power of the mode equals $|A_0|^2$:
\begin{align}
\begin{split}
\iint_{A_\infty}\frac{1}{2}\Re \{|A_0|^2 \frac{\mathbf{e}(\omega_j)}{\sqrt{\mathscr{P}_j}}\times \frac{\mathbf{h}^*(\omega_j)}{\sqrt{\mathscr{P}_j}} \} & \cdot \mathbf{\hat{e}}_z dA \equiv |A_0|^2 \\ \Rightarrow \mathscr{P}_j &= \frac{1}{4} \iint_{A_\infty}\{ \mathbf{e}(\omega_j) \times\mathbf{h}^*(\omega_j) + \mathbf{e}^*(\omega_j) \times\mathbf{h}(\omega_j))  \}\cdot \mathbf{\hat{e}}_z dA \;. \label{eq:normalization}
\end{split}
\end{align}
$A_\infty$ is the plane perpendicular to the waveguide propagation direction. $\mathbf{\hat{e}}_z$ is the unit vector in the propagation direction $z$. By definition, these modes obey the Maxwell curl equations,
\begin{align}
\nabla \times \mathbf{E}_{0}(\omega_j)=& i\omega_j \mu_0 \mathbf{H}_{0}(\omega_j)\;,\label{eq:MaxwellCurlE}\\
\nabla \times \mathbf{H}_{0}(\omega_j)=& -i\omega_j \epsilon_0n^2 \mathbf{E}_{0}(\omega_j)\;,\label{eq:MaxwellCurlH}
\end{align}
$n(\mathbf{r}_\perp)$ is the index of the unperturbed waveguide cross-section. One can include the effect of perturbations, such as linear losses and nonlinearities, by introducing complex slowly varying amplitudes $A_j(z)$. The perturbed waveguide modes are then written as: 
\begin{align}
\mathbf{E}(\omega_j, \mathbf{r})=& A_j(z) \frac{\mathbf{e}(\omega_j,\mathbf{r}_\perp)}{\sqrt{\mathscr{P}_j}}e^{i\beta_j z} \;,\label{eq:EPerturbed}\\
\mathbf{H}(\omega_j,\mathbf{r})=& A_j(z) \frac{\mathbf{h}(\omega_j,\mathbf{r}_\perp)}{\sqrt{\mathscr{P}_j}}e^{i\beta_j z} \;.\label{eq:HPerturbed}
\end{align}
In practice, we will consider the total field to be a superposition of a number of monochromatic waves:
\begin{align}
\widetilde{\mathbf{E}}(\mathbf{r},t)=& \sum_j\Re\{ A_j(z) \frac{\mathbf{e}(\omega_j, \mathbf{r}_\perp)}{\sqrt{\mathscr{P}_j}}e^{-i(\omega_j t - \beta_j z)}\} \;,\\
\widetilde{\mathbf{H}}(\mathbf{r},t)=& \sum_j\Re\{ A_j(z) \frac{\mathbf{h}(\omega_j, \mathbf{r}_\perp)}{\sqrt{\mathscr{P}_j}}e^{-i(\omega_j t - \beta_j z)}\} \;.
\end{align}
These perturbed modes should also obey the Maxwell curl equations, where the influence of the graphene sheet can be incorporated as a current density, $\mathbf{J}(\omega_j)$,
\begin{align}
\nabla \times \mathbf{E}(\omega_j)=& i\omega_j \mu_0 \mathbf{H}(\omega_j)\;,\label{eq:MaxwellCurlEpert}\\
\nabla \times \mathbf{H}(\omega_j)=& -i\omega_j \epsilon_0n^2 \mathbf{E}(\omega_j) +\mathbf{J}(\omega_j)\;.\label{eq:MaxwellCurlHpert}
\end{align}
This current density can be written as the sum of a linear and a nonlinear contribution,
\begin{align}
\begin{split}
\mathbf{J}(\omega_j) =&\mathbf{J}_{L}(\omega_j)+\mathbf{J}_{NL}(\omega_j)\\
=& \sigma^{(1)}(\omega_j) \mathbf{E}(\omega_j) + \frac{1}{4}\sum_{\omega_j=\omega_k+\omega_l+\omega_m} \sigma^{(3)}(\omega_j; \omega_k, \omega_l, \omega_m)\vdots\mathbf{E}(\omega_k)\mathbf{E}(\omega_l) \mathbf{E}(\omega_m)\;,
\end{split}\label{eq:CurrentDensity}
\end{align}
$\sigma^{(1)}$ and $\sigma^{(3)}$ are the first and third order conductivity tensors. To derive the coupled-wave equations, we can start from the conjugated form of the Lorentz reciprocity theorem \cite{S_osgood2009engineering}:
\begin{align}
\iint_{A_\infty} \nabla \cdot \mathbf{F} = \frac{\partial}{\partial z} \iint_{A_\infty} \mathbf{F} \cdot \mathbf{\hat{e}}_z dA \;.
\label{eq:divtheo}
\end{align}
$A_\infty$ is the surface perpendicular to the propagation direction. The $\mathbf{F}$-field can be constructed from the perturbed and unperturbed waveguide mode fields as $ \mathbf{F} \equiv \mathbf{E}_{0}^*(\omega_j)\times \mathbf{H}(\omega_j) + \mathbf{E}(\omega_j)\times \mathbf{H}_{0}^*(\omega_j)$. Substituting this in Eq. \eqref{eq:divtheo} yields:
\begin{align}
\begin{split}
\iint_{A_\infty} \{ (\nabla \times \mathbf{E}_{0}^*(\omega_j))\cdot \mathbf{H}(\omega_j) &- \mathbf{E}_{0}^*(\omega_j) \cdot (\nabla \times  \mathbf{H}(\omega_j)) \\ 
+ (\nabla \times \mathbf{E}(\omega_j))\cdot \mathbf{H}_{0}^*(\omega_j) &- \mathbf{E}(\omega_j) \cdot (\nabla \times  \mathbf{H}_{0}^*(\omega_j))\}  dA\\
&= \frac{\partial}{\partial z}\iint_{A_\infty}\frac{A_0^* A_j(z)}{\mathscr{P}_j} \{ \mathbf{e}(\omega_j) \times\mathbf{h}^*(\omega_j) + \mathbf{e}^*(\omega_j) \times\mathbf{h}(\omega_j))  \}\cdot \mathbf{\hat{e}}_z dA \;.
\end{split}
\label{eq:divtheo_subs}
\end{align}
The left hand side of Eq. \eqref{eq:divtheo_subs} can be simplified by substituting Eqs. \eqref{eq:MaxwellCurlE}-\eqref{eq:MaxwellCurlH} and \eqref{eq:MaxwellCurlEpert}-\eqref{eq:MaxwellCurlHpert}. The right hand side can be simplified by using the normalization condition (Eq. \eqref{eq:normalization}). Eventually this gives
\begin{align}
\frac{\partial}{\partial z}A_j= -\frac{e^{-i\beta_j z}}{4\sqrt{\mathscr{P}_j}} \iint_{A_\infty} \mathbf{e}^*(\omega_j)\cdot\mathbf{J}(\omega_j) dA\;,
\label{eq:divtheo_subs_final}
\end{align}
substituting Eqs. \eqref{eq:EPerturbed} and \eqref{eq:HPerturbed} in Eq. \eqref{eq:CurrentDensity}, and subsequently in Eq. \eqref{eq:divtheo_subs_final}, one gets a general coupled-wave equation for the set of slowly varying amplitudes:
\begin{align}
\begin{split}
\frac{\partial}{\partial z}A_j=&-\frac{A_j}{4\mathscr{P}_j}\iint_{A_\infty} \mathbf{e}^*(\omega_j) \cdot \sigma^{(1)}(\omega_j)\mathbf{e}(\omega_j) dA \\ -\sum_{\substack{\omega_j=\omega_k \\ +\omega_l+\omega_m}} & \frac{A_k A_l A_m e^{i(\beta_k+\beta_l+\beta_m-\beta_j)z}}{16\sqrt{\mathscr{P}_j\mathscr{P}_k\mathscr{P}_l\mathscr{P}_m}}\iint_{A_\infty}\mathbf{e}^*(\omega_j)\cdot\sigma^{(3)}(\omega_j; \omega_k, \omega_l, \omega_m)\vdots\mathbf{e}(\omega_k)\mathbf{e}(\omega_l)\mathbf{e}(\omega_m)dA\;, \label{eq:CoupledWaveEqs_general}
\end{split}
\end{align}
here the summation goes over all possible combinations of 3 frequencies that add up to $\omega_j$, including the negative frequencies. Moreover, since the time-dependent electrical fields are real-valued, one can make use of the equality $\mathbf{e}(-\omega)=\mathbf{e}^*(\omega)$.

In the case of degenerate four-wave mixing, there are 3 monochromatic waves involved, the pump, signal and idler, at equally spaced frequency intervals ($\Delta\omega=\omega_s-\omega_p=\omega_p-\omega_i$). For this specific case, the coupled wave equations (Eq. \eqref{eq:CoupledWaveEqs_general}) can be simplified to:
\begin{align}
\begin{split}
\frac{\partial A_p}{\partial z}&=i\{\gamma(\omega_p;\omega_p,\omega_p,-\omega_p)\vert A_p\vert^2 A_p + 2\gamma(\omega_p;\omega_p,\omega_s,-\omega_s)\vert A_s\vert^2 A_p \\
&+ 2\gamma(\omega_p;\omega_p,\omega_i,-\omega_i)\vert A_i\vert^2 A_p  + 2\gamma(\omega_p;\omega_s,\omega_i,-\omega_p) A_sA_iA_p^* e^{-i\Delta\beta z}\}-\frac{\alpha(\omega_p)}{2}A_p \;,\label{eq:CoupledWaveEqs_degenerateFWM1}
\end{split}\\
\begin{split}
\frac{\partial A_s}{\partial z}&=i\frac{\omega_s}{\omega_p}\{\gamma(\omega_s;\omega_s,\omega_s,-\omega_s)\vert A_s\vert^2 A_s + 2\gamma(\omega_s;\omega_s,\omega_p,-\omega_p)\vert A_p\vert^2 A_s \\
&+ 2\gamma(\omega_s;\omega_s,\omega_i,-\omega_i)\vert A_i\vert^2 A_s  + \gamma(\omega_s;\omega_p,\omega_p,-\omega_i) A_pA_pA_i^* e^{i\Delta\beta z}\}-\frac{\alpha(\omega_s)}{2}A_s \;,\label{eq:CoupledWaveEqs_degenerateFWM2}
\end{split}\\
\begin{split}
\frac{\partial A_i}{\partial z}&=i\frac{\omega_i}{\omega_p}\{\gamma(\omega_i;\omega_i,\omega_i,-\omega_i)\vert A_i\vert^2 A_i + 2\gamma(\omega_i;\omega_i,\omega_p,-\omega_p)\vert A_p\vert^2 A_i \\
&+ 2\gamma(\omega_i;\omega_i,\omega_s,-\omega_s)\vert A_s\vert^2 A_i  + \gamma(\omega_i;\omega_p,\omega_p,-\omega_s) A_pA_pA_s^* e^{i\Delta\beta z}\}-\frac{\alpha(\omega_i)}{2}A_i \;,\label{eq:CoupledWaveEqs_degenerateFWM3}
\end{split}
\end{align}
where $A_p(z)$, $A_s(z)$ and $A_i(z)$ are the complex amplitudes of respectively the pump, signal and idler, normalized so that $\vert A_{p,s,i}\vert^2$ equals the total power in the respective mode. $\Delta\beta = 2\beta(\omega_p)-\beta(\omega_s)-\beta(\omega_i)$ is the phase mismatch term. $\alpha(\omega)$ represents the linear loss and $\gamma(\omega_p+\omega_q+\omega_r;\omega_p,\omega_q,\omega_r)$ is the nonlinear parameter of the waveguide. Comparing with Eq. \eqref{eq:CoupledWaveEqs_general}, these parameters become:
\begin{align}
\begin{split}
\alpha(\omega_j)=&\frac{1}{2\mathscr{P}_j}\iint_{A_\infty} \mathbf{e}^*(\omega_j) \cdot \sigma^{(1)}(\omega_j)\mathbf{e}(\omega_j) dA\;,
\end{split}\\
\begin{split}
\gamma(\omega_j=\omega_p+\omega_q+\omega_r;\omega_p,\omega_q,\omega_r)=& \\ i\frac{3}{N_{(p,q,r)}} \sum_{k,l,m}\frac{1}{16\sqrt{\mathscr{P}_j\mathscr{P}_k\mathscr{P}_l\mathscr{P}_m}}&\iint_{A_\infty}\mathbf{e}^*(\omega_j)\cdot\sigma^{(3)}(\omega_j; \omega_k, \omega_l, \omega_m)\vdots\mathbf{e}(\omega_k)\mathbf{e}(\omega_l)\mathbf{e}(\omega_m)dA\;.\label{eq:gamma_general}
\end{split}
\end{align}
In Eq. \eqref{eq:gamma_general}, the summation parameters $(k,l,m)$ take all different permutations of the set $(p,q,r)$. $N_{(p,q,r)}$ is the number of permutations of the set $(p,q,r)$. In the specific case of graphene, $\sigma^{(1)}$ and $\sigma^{(3)}$ are only present in a very thin layer. The effects can be very well described using first and third order \textit{sheet} conductivities, $\sigma^{(1)}_s$ and $\sigma^{(3)}_s$. The surface integrals then become line integrals over the graphene,
\begin{align}
\begin{split}
\alpha(\omega_j)=&\frac{1}{2\mathscr{P}_j}\int_{G} \mathbf{e}^*(\omega_j) \cdot \sigma^{(1)}_s(\omega_j)\mathbf{e}(\omega_j) d\ell\;, \label{eq:alpha_general2D}
\end{split}\\
\begin{split}
\gamma(\omega_j=\omega_p+\omega_q+\omega_r;\omega_p,\omega_q,\omega_r)=& \\ i \frac{3}{N_{(p,q,r)}} \sum_{k,l,m}\frac{1}{16\sqrt{\mathscr{P}_j\mathscr{P}_k\mathscr{P}_l\mathscr{P}_m}}&\int_{G}\mathbf{e}^*(\omega_j)\cdot\sigma^{(3)}_s(\omega_j; \omega_k, \omega_l, \omega_m)\vdots\mathbf{e}(\omega_k)\mathbf{e}(\omega_l)\mathbf{e}(\omega_m)d\ell\;.\label{eq:gamma_general2D}
\end{split}
\end{align}
In the specific case of the four-wave mixing experiment described in this work, the coupled-wave equations in Eqs. \eqref{eq:CoupledWaveEqs_degenerateFWM1}-\eqref{eq:CoupledWaveEqs_degenerateFWM3} can be strongly simplified. Firstly, the pump carries a much higher power than the signal, moreover the idler will be orders of magnitude weaker ($\vert A_p \vert > \vert A_s \vert \gg \vert A_i \vert$). Secondly, in Ref. \cite{S_alexander2015electrically} it was demonstrated for a similar waveguide platform that nonlinear absorption only starts affecting the overall power transmission significantly for power levels above 1 W. In the experiments presented here the on-chip power levels were kept on the order of 10 mW or lower. Hence self-phase/amplitude modulation is expected to be much weaker than linear absorption ($\vert\gamma\vert\vert A_p \vert^2 \ll \vert\frac{\alpha(\omega)}{2}\vert$). Thirdly, the phase mismatch is negligible ($L\beta_2\Delta\omega^2\ll 1$, $L=100\;\mu$m, $\Delta\omega<10^{13}$ rad/s and $\beta_2\equiv\frac{\partial^2 \beta}{\partial \omega^2}$ of a SiN waveguide is on the order of $10^{-25}$ s$^2$/m \cite{moss2013new}). All these assumptions lead to heavily simplified coupled-wave equations:
\begin{align}
\frac{\partial A_p}{\partial z}\approx&-\frac{\alpha(\omega_p)}{2}A_p \;,\\
\frac{\partial A_s}{\partial z}\approx&-\frac{\alpha(\omega_s)}{2}A_s \;,\\
\frac{\partial A_i}{\partial z}\approx& i\frac{\omega_i}{\omega_p}\gamma(\omega_i;\omega_p,\omega_p,-\omega_s) A_pA_pA_s^* -\frac{\alpha(\omega_i)}{2}A_i \;.
\end{align}
Under these conditions the conversion efficiency $\eta$, defined as the ratio of the idler power to the signal power, has a quadratic dependence on the nonlinear parameter $\gamma$ \cite{S_dave2015nonlinear}:
\begin{align}
\begin{split}
\eta\equiv\frac{P_i(L)}{P_s(L)} & =\frac{\vert A_i(L)\vert^2}{\vert A_s(L)\vert^2}\\
& \approx\vert \gamma(\omega_i;\omega_p,\omega_p,-\omega_s) \vert^2 \left(\frac{\omega_i}{\omega_p} \right)^2 P_p(0)^2L_{\text{eff}}^2 e^{\{\alpha(\omega_s)-\alpha(\omega_i)\}L}\\
& \approx\vert \gamma(\omega_i;\omega_p,\omega_p,-\omega_s) \vert^2 \left(\frac{\omega_i}{\omega_p} \right)^2 P_p(0)^2L_{\text{eff}}^2\label{eq:eta}\;,
\end{split}
\end{align}
where the effective interaction length is defined as:
\begin{align}
L_{\text{eff}}\equiv\frac{1-e^{-\{\alpha(\omega_p)+\alpha(\omega_s)/2-\alpha(\omega_i)/2\} L}}{\alpha(\omega_p)+\alpha(\omega_s)/2-\alpha(\omega_i)/2}\approx \frac{1-e^{-\alpha L}}{\alpha}\label{eq:Leff}\;.
\end{align}
The final expressions in Eqs. \eqref{eq:eta} and \eqref{eq:Leff} are only valid when the approximation $\alpha(\omega_p)\approx\alpha(\omega_s)\approx\alpha(\omega_i)\equiv \alpha$ holds, i.e. when the frequencies detuning is small ($\Delta \omega \ll \omega_p)$. For the specific experiments described in this work, the expressions for $ \alpha(\omega)$ and $\gamma(\omega_i;\omega_p,\omega_p,-\omega_s)$ can be further simplified. It is assumed that a flat sheet of graphene lies in the xz plane (this can easily be generalized to arbitrary graphene shapes). Firstly, the linear conductivity has only two nonzero elements, which are equal: $\sigma^{(1)}_{xx}=\sigma^{(1)}_{zz}$. Now we can treat the linear conductivity as a scalar parameter and calculate the linear loss as:
\begin{align}
\alpha(\omega_j)=\frac{\sigma^{(1)}_{s,xx}(\omega_j)}{2\mathscr{P}_j}\int_{G} \vert\mathbf{e}(\omega_j)_\parallel\vert^2d\ell\;,
\end{align}
where $\mathbf{e}(\omega_j)_\parallel$ is the electric field component tangential to the graphene sheet. Using symmetry considerations one can prove that the third order conductivity tensor of graphene has only the following nonzero elements \cite{S_cheng2014}:
\begin{align}
\sigma^{(3)}_{s,\;xxxx}&=\sigma^{(3)}_{s,\;zzzz}\;,\\
\sigma^{(3)}_{s,\;xxzz}&=\sigma^{(3)}_{s,\;zzxx}\;,\\
\sigma^{(3)}_{s,\;xzxz}&=\sigma^{(3)}_{s,\;zxzx}\;,\\
\sigma^{(3)}_{s,\;xzzx}&=\sigma^{(3)}_{s,\;zxxz}\;,\\
\sigma^{(3)}_{s,\;xxxx}&=\sigma^{(3)}_{s,\;xxzz}+\sigma^{(3)}_{s,\;xzxz}+\sigma^{(3)}_{s,\;xzzx}\;.
\end{align}
Moreover, simulations show that the modes in the SiN waveguides used in this work are quasi-transversal, meaning that $\mathrm{e}_x\gg\mathrm{e}_z$. This implies that the term containing $\sigma^{(3)}_{s,\;xxxx}$ in Eq. \eqref{eq:gamma_general2D} is about two orders of magnitude larger than any of all other terms, the expression for the nonlinear parameter can be simplified to:
\begin{align}
\gamma(\omega_i;\omega_p,\omega_p,-\omega_s) &\approx i  \frac{3 \sigma^{(3)}_{s,\;xxxx}(\omega_i; \omega_p, \omega_p, -\omega_s)}{16\mathscr{P}_p\sqrt{\mathscr{P}_i\mathscr{P}_s}}\int_{G}\mathrm{e}^*(\omega_i)_x\mathrm{e}(\omega_p)_x\mathrm{e}(\omega_p)_x\mathrm{e}^*(\omega_s)_x  d\ell \\
&\approx i \frac{3 \sigma^{(3)}_{s,\;xxxx}(\omega_i; \omega_p, \omega_p, -\omega_s)}{16\mathscr{P}_p^2}\int_{G}\vert\mathrm{e}(\omega_p)_x\vert^4  d\ell\;.
\end{align}
To arrive to the second expression, we have used the assumption that $\mathrm{e}(\omega_p)\approx\mathrm{e}(\omega_s)\approx\mathrm{e}(\omega_i)$, which is the case when one considers the same spatial modes and small detunings ($\Delta\omega \ll \omega_p$). We can further generalize this expression to arbitrary graphene shapes:
\begin{align}
\gamma(\omega_i;\omega_p,\omega_p,-\omega_s)
&\approx i \frac{3 \sigma^{(3)}_{s,\;xxxx}(\omega_i; \omega_p, \omega_p, -\omega_s)}{16\mathscr{P}_p^2}\int_{G}\vert\mathbf{e}(\omega_p)_\parallel\times \mathbf{\hat{e}}_z\vert^4  d\ell\;.
\end{align}
In this Letter, the electric field profile $\mathbf{e}$ was calculated for the cross-section of the waveguide using COMSOL Multiphysics$^\circledR$. The above formula was then used to convert between the waveguide nonlinear parameter $\gamma$ and the material parameter $\sigma^{(3)}_{s}$.
\section{Graphene gating using polymer electrolyte}
\begin{figure}
    \includegraphics[width=0.9\textwidth]{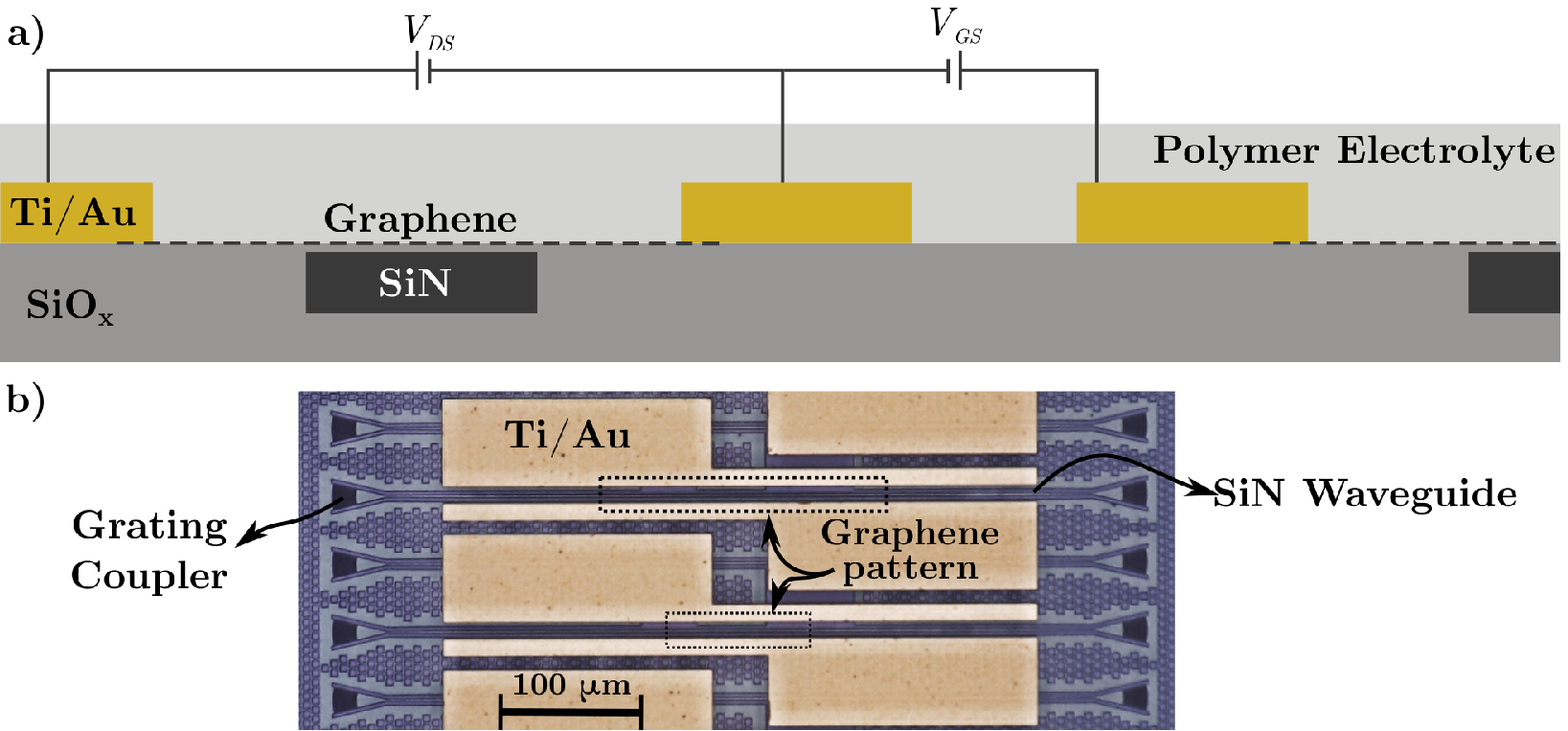}
    \caption{\label{fig:gating_micrograph}\normalsize a) Sketch of the sample cross-section and gating scheme (not to scale). b) Optical microscope image of a set of waveguides. The SiN waveguides  can be seen clearly, as well as the grating couplers. Every other waveguide is covered with a section of graphene. Graphene is not visible on the optical microscope image but the extent of the graphene (under the contacts) is shown by the dashed lines. The graphene is contacted at both sides. On top of this structure the polymer electrolyte is spin-coated (not in this image).}
\end{figure}
\begin{figure}[b]	
	\includegraphics{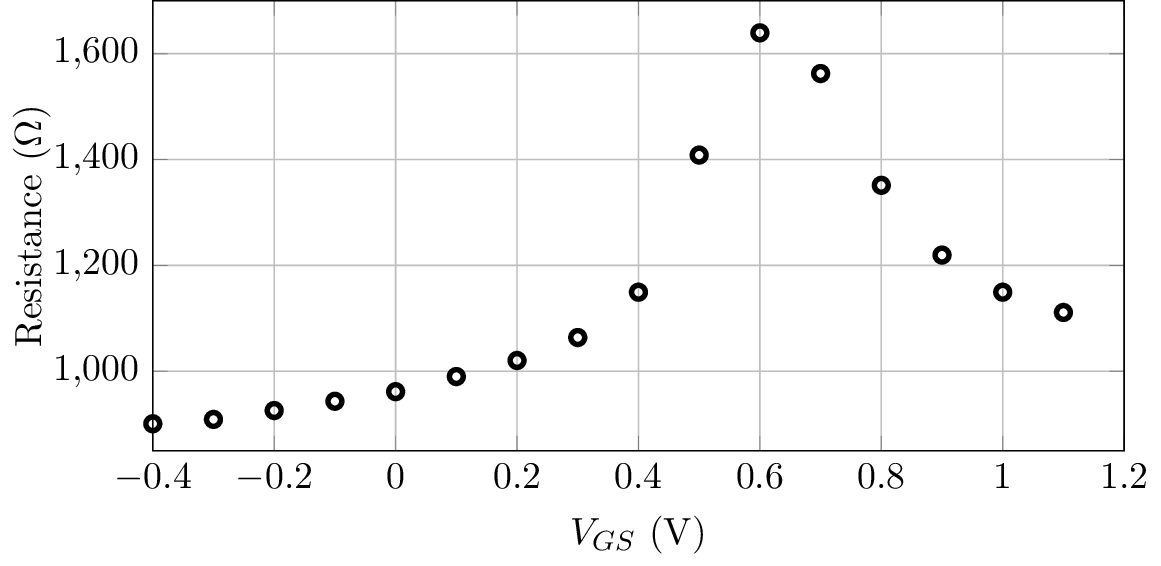}
		\caption{\normalsize The measured electrical resistance over the graphene as function of the gate voltage $V_{GS}$.\label{fig:measurement_resistance}} 
\end{figure}
\begin{figure}[t]
	\includegraphics{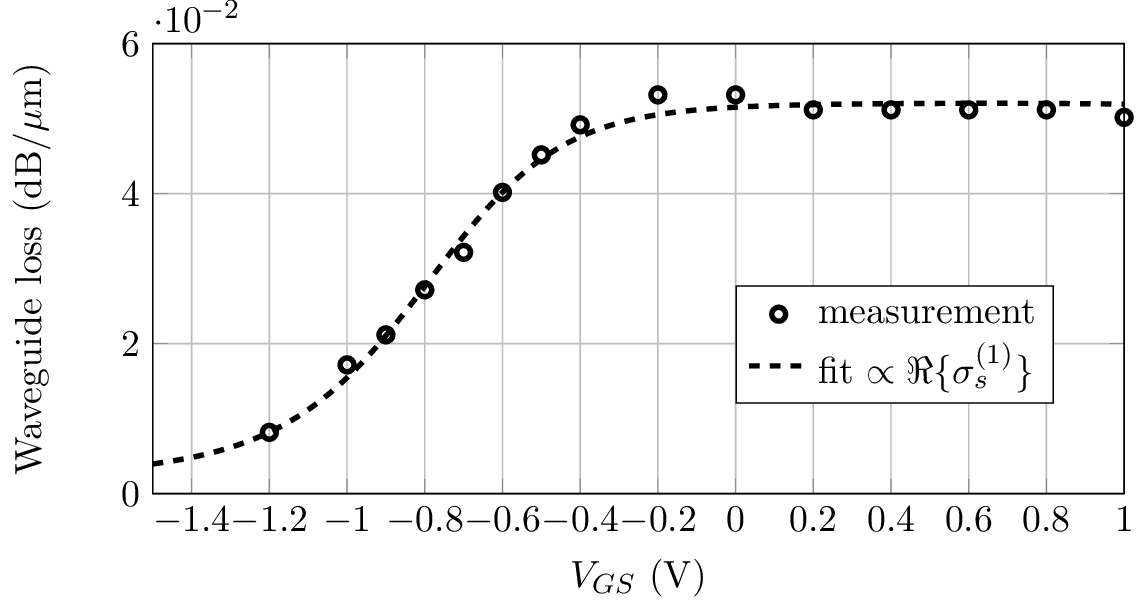}
		\caption{\normalsize The measured optical loss and corresponding fit, proportional to the real part of the linear conductivity of graphene $\sigma^{(1)}_s$.\label{fig:measurement_loss}} 
\end{figure}
\begin{figure}[b]	
	\includegraphics{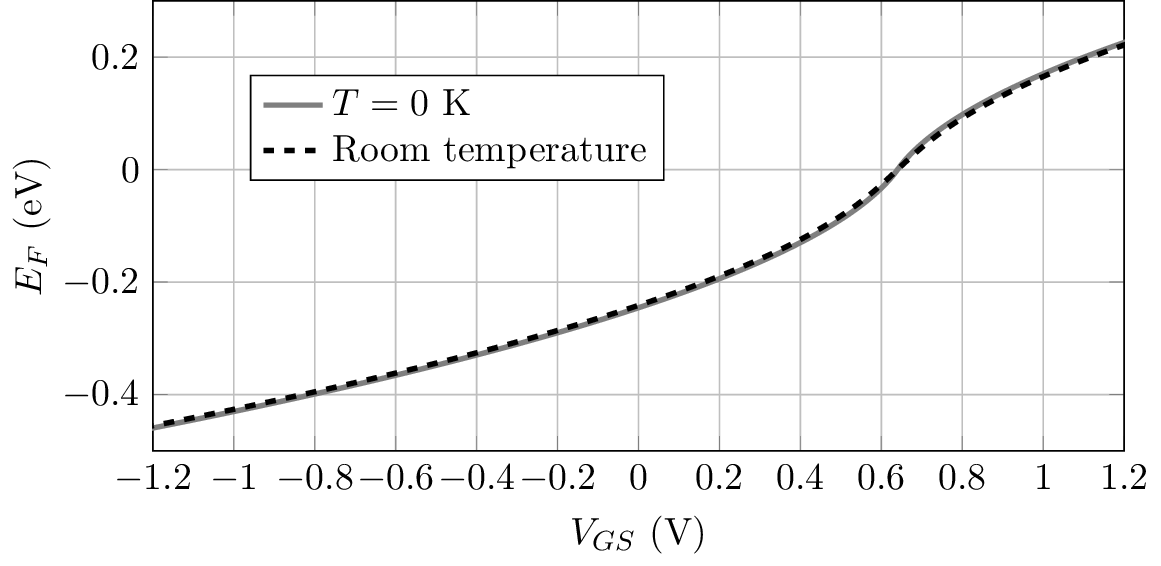}
		\caption{\normalsize Estimated relation between gate voltage $V_{GS}$ and Fermi energy $E_F$ of the graphene covering the waveguide used in this work.\label{fig:Ef_vs_V}} 
\end{figure}
In this work, a polymer electrolyte (LiClO$_4$ and polyethylene oxide in a weight ratio of 0.1:1) is used to gate the graphene, i.e. to electrostatically change the carrier density or Fermi energy of the graphene. Fig. \ref{fig:gating_micrograph}a shows a sketch of the sample cross-section. Each waveguide is covered by patterned graphene, which is contacted at both sides, making a simple resistance measurement possible by applying a voltage $V_{DS}$. The whole sample is covered with the polymer electrolyte, by applying a voltage $V_{GS}$ to a `gate' contact (in principle any isolated contact in the vicinity of the waveguide, in the Letter the contact on the adjacent waveguide is used) the carrier density in the graphene can be tuned. The dependence of the Fermi energy $E_F$ on the gate voltage $V_{GS}$ can be approximated by the following formula \cite{S_thareja2015electrically, S_das2008monitoring}:
\begin{align}
V_{GS}-V_D = \mathrm{sgn}(E_F)\frac{e E_F^2}{\hbar^2 v_F^2 \pi C_{EDL}}+\frac{E_F}{e} \label{eq:VvsEf}\;,
\end{align}
with $e$ the electron charge, $v_F\approx 10^6$ m/s the Fermi velocity and $C_{EDL}$ the electric double layer capacitance. $V_D$ is the Dirac voltage, the voltage at which the graphene becomes intrinsic and at which the conductance reaches a minimum. Note that Eq. \eqref{eq:VvsEf} is derived at temperature T=0 K, but the difference with room temperature is negligibly small (see Fig. \ref{fig:Ef_vs_V}). To obtain an estimate of $V_D$, an electrical resistance measurement of the gated graphene is used. Fig. \ref{fig:measurement_resistance} shows the total resistance between the source and drain electrode as a function of the $V_{GS}$ voltage. Based on this measurement we estimate $V_D\approx 0.64$ V. To estimate the capacitance $C_{EDL}$, a measurement of the optical loss through the waveguide, as a function of $V_{GS}$ can be used. Fig. \ref{fig:measurement_loss} shows the measured loss and corresponding fit, which is proportional to the real part of the linear conductivity $\sigma^{(1)}_s$, which was calculated using the Kubo formula \cite{S_falkovsky2007space, S_mikhailov2007, S_falkovsky2007optical}. The obtained value for $C_{EDL}$ is $1.8\cdot10^{-2}$ F m$^{-2}$ (other used parameters for this fit were $\hbar\Gamma=10$ meV and $T=293$ K). The resulting relation between Fermi energy and gate voltage is shown in Fig. \ref{fig:Ef_vs_V}.

Figure \ref{fig:gating_micrograph}b shows an optical microscope image of an actual set of contacted graphene-covered waveguides. The SiN waveguides can be seen, as well as the grating couplers used to couple to the optical fiber. To provide enough space for the contact needles, only every other waveguide is covered with a section of graphene. The graphene is not visible on the optical microscope image, therefore its extent is shown by the dashed lines. The graphene is contacted at both sides. On top of this structure the polymer electrolyte is spin-coated (not in this image).

\section{Evaluation of $\sigma_{s,xxxx}^{(3)}$}
\begin{figure}[h]
	\includegraphics{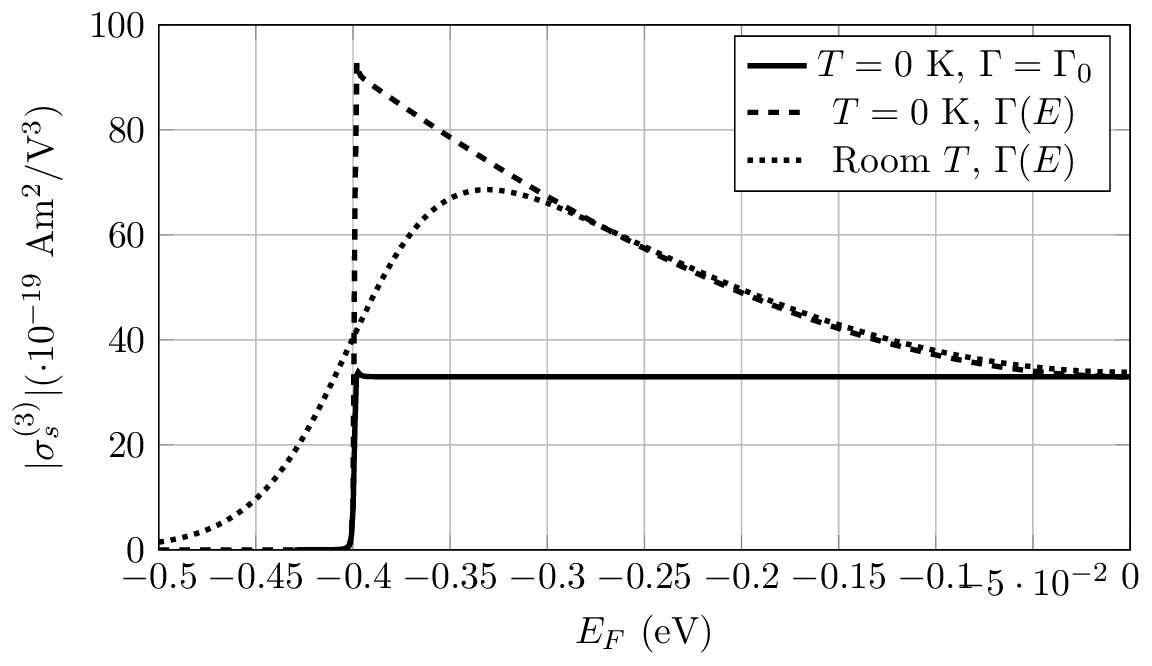}
		\caption{\label{fig:diff_models} \normalsize The absolute value of the third order conductivity $|\sigma^{(3)}_{s,xxxx}(\omega_p;\omega_p,\omega_p,-\omega_p)|$ as a function of Fermi energy at $\lambda_s=\lambda_p=1550.18$ nm. Solid curve: $T=0$, $\hbar\Gamma=\hbar\Gamma_0=2.5$ meV; dashed ($T=0$) and dotted ($k T=25$ meV) curves are plotted at $\hbar\Gamma(E)$ determined by Eq. (4) of the main text with $\Gamma_0 = 2.5$ meV, $E_0 = 250$ meV and $\alpha = 0.8$.}
\end{figure}
We compare our experimental results with the theory of Refs. \cite{S_Cheng2015, S_Mikhailov2016}. Analytical expressions for the third order conductivity $\sigma^{(3)}_{s,\alpha\beta\gamma\delta}(\omega_i;\omega_p,\omega_p,-\omega_s)$ have been derived in these papers at $T=0$ and at the relaxation rate $\Gamma=\Gamma_0$ independent of energy. The calculated $|\sigma^{(3)}_{s,xxxx}(\omega_p;\omega_p,\omega_p,-\omega_p)|$ is shown in Fig. \ref{fig:diff_models} by the solid curve; one can see that the Fermi-energy dependence of the third order conductivity at $T=0$ and $\Gamma=\Gamma_0$ has a step-like shape (compare with Fig. 9(b) from Ref. \cite{S_Mikhailov2016}) which does not look similar to the experimental curves in Fig. 3(a) of the main text. However, as discussed in the main text, a more realistic model of the relaxation rate supposes an energy dependence of $\Gamma(E)$ given by Eq. (4) of the main Letter. Using this model for $\Gamma(E)$ we plot $|\sigma^{(3)}_{s,xxxx}(\omega_p;\omega_p,\omega_p, -\omega_p)|$ by the dashed curve at $T=0$ and by the dotted curve at room temperature ($kT=25$ meV); a good qualitative agreement of the dotted curve with the experimental data is now evident. The parameters $\Gamma_0$, $\alpha$ and $E_0$ are chosen to quantitatively fit the characteristic features of the curves $|\sigma^{(3)}(E_F)|$ and $|\sigma^{(3)}(\lambda_s-\lambda_p)|$ (e.g. the linewidth, the maxima and minima) to the experimental ones.

\def\bibfont{\normalsize}

\end{document}